\documentclass[modern]{aastex62}

\usepackage{graphicx,epsfig,natbib,color}
\usepackage{epstopdf}
\usepackage{subfigure}
\usepackage{graphicx}
\usepackage{epsfig}
\usepackage{natbib}

\shorttitle{Investigating the Formation and Eruption}
\shortauthors{Huang et al.}
\begin{document}

\title{\textbf{Formation and Eruption of a Mini-sigmoid Originating in Coronal Hole}}

\correspondingauthor{}
\email{Huangzw@smail.nju.edu.cn; xincheng@nju.edu.cn}

\author{Z. W. Huang}
\affil{School of Astronomy and Space Science, Nanjing University, Nanjing 210023, China\\}
\affil{Key Laboratory of Modern Astronomy and Astrophysics (Nanjing University), Ministry of Education, Nanjing 210093, China\\}
\author{X. Cheng}
\affil{School of Astronomy and Space Science, Nanjing University, Nanjing 210023, China\\}
\affil{Key Laboratory of Modern Astronomy and Astrophysics (Nanjing University), Ministry of Education, Nanjing 210093, China\\}
\author{Y. N. Su}
\affil{Key Laboratory for Dark Matter and Space Science, Purple Mountain Observatory, CAS, Nanjing 210008, China\\}
\affil{School of Astronomy and Space Science, University of Science and Technology of China, Hefei, Anhui 230026, People’s Republic of China\\}
\author{T. Liu}
\affil{Key Laboratory for Dark Matter and Space Science, Purple Mountain Observatory, CAS, Nanjing 210008, China\\}
\affil{School of Astronomy and Space Science, University of Science and Technology of China, Hefei, Anhui 230026, People’s Republic of China\\}
\author{M. D. Ding}
\affil{School of Astronomy and Space Science, Nanjing University, Nanjing 210023, China\\}
\affil{Key Laboratory of Modern Astronomy and Astrophysics (Nanjing University), Ministry of Education, Nanjing 210093, China\\}

\begin{abstract}

In this paper, we study in detail the evolution of a mini-sigmiod originating in a cross-equatorial coronal hole, where the magnetic field is mostly open and seriously distinct from the closed background field above active-region sigmoids. The source region first appeared as a bipole, which subsequently experienced a rapid emergence followed by a long-term decay. Correspondingly, the coronal structure initially appeared as arc-like loops, then gradually sheared and transformed into continuously sigmoidal loops, mainly owing to flux cancellation near the polarity inversion line. The temperature of J-shaped and sigmoidal loops is estimated to be about $2.0\times10^{6}$ K, greater than that of the background coronal hole. Using the flux-rope insertion method, we further reconstruct the nonlinear force-free fields that well reproduces the transformation of the potential field into a sigmoidal field. The fact that the sheared and sigmoidal loops are mainly concentrated at around the high-Q region implies that the reconnection most likely takes place there to form the sigmoidal field and heat the plasma. Moreover, the twist of sigmoidal field lines is estimated to be around 0.8, less than the values derived for the sigmoids from active regions. However, the sigmoidal flux may quickly enter an unstable regime at the very low corona ($<10\ Mm$) due to the open background field. The results suggest that the mini-sigmoid, at least the one in our study, has the same formation and eruption process as the large-scale one, but is significantly influenced by the overlying flux.

\end{abstract}


\section{Introduction}

Solar eruptions refer to the phenomena involving explosive outflows of magnetic field and plasma from the solar atmosphere. Coronal mass ejections (CMEs) are the most violent solar eruptions and usually accompany with solar flares \citep{2018SSRv..214...46G,2003SoPh..218..261A,2005JGRA..11012S05Y}. Before the eruption of CMEs, S-shaped structures, i.e., sigmoids \citep{1996ApJ...464L.199R}, are often observed in soft X-rays in their source regions \citep{1998GeoRL..25.2481H,1996ApJ...464L.199R,2002ApJ...574.1021G,2007ApJ...669.1372L,2009ApJ...700L..83G}. Using full-resolution images from the Yohkoh Soft X-Ray Telescope, \citet{1999GeoRL..26..627C,2007ApJ...671L..81C} studied 107 bright sigmoids and found that the active regions with sigmoids are more likely to produce large-scale eruptions. A relationship is also found between the chirality and the location of sigmoids, i.e., most of sigmoids in the southern hemisphere are right-handed, manifesting as a forward S-shape and those in the north are left-handed, appearing as a reversed S-shape \citep{1996ApJ...464L.199R,1997SoPh..175...27Z}. Based on their lifetime, sigmoids are also classified as transient ones, which generally last for tens of minutes to hours and always present a single and continuous loop, and long-lasting ones, which last for days to weeks and generally comprises of stable multiple loops \citep{2009ApJ...691.1276A,2001ApJ...552..833M}. Note that, it is found that two discontinuous J-shaped structures, connecting the front (back) of the positive polarity to the front (back) of the negative polarity, are also displayed as a single sigmoid apparently \citep{1996ApJ...473..533P}. 

It is usually considered that sigmoids are kind of observational manifestations of a flux rope \citep{1998GeoRL..25.2481H,1996ApJ...464L.199R,1997ApJ...491L..55S,2017ScChE..60.1383C}, which is a specific magnetic structure composed of a bundle of magnetic field lines twisting around a common axis. As for the formation of the flux rope, two groups of models have been proposed. The first group suggests that the flux rope preexists in the convection zone and then emerges into the corona by buoyancy \citep{2001ApJ...554L.111F,2004ApJ...609.1123F,2006ApJ...641L.149F,2006JGRA..11112103G,2008A&A...492L..35A,2009ApJ...691.1276A}. However, \citet{2001ApJ...554L.111F} found that the flux rope is unable to completely pass across the photospheric boundary due to the drag of plasma gathered in magnetic dips. The flux rope in the corona is actually a newly formed one through the reconnection of the partially emerged field (also see \citealt{2009A&A...503..999H,2009A&A...507..995M}). The second mechanism suggests that the flux rope is built up directly via the reconnection low in the solar atmosphere, which transforms sheared arcades into twisted field lines. Meanwhile, some small loops are formed and then submerge under the photosphere due to magnetic tension, manifesting as flux cancellation \citep{1989ApJ...343..971V,2011A&A...526A...2G, Gibb_2014}. Based on shearing and converging motions, as well as magnetic cancellation near the polarity inversion line (PIL), several numerical models have successfully reproduced the transformation of potential field lines to long-lasting sigmoidal ones \citep{2000ApJ...529L..49A,2003ApJ...585.1073A,2005A&A...430.1067A,2010ApJ...708..314A,2001ApJ...560..445M,2005ApJ...621L..77M,2006ApJ...641..577M}. \citet{Liu_2002} also found that the reconnection may transfer the mutual helicity into the self-helicity to form the sigmoidal structure.

Observationally, many investigations also addressed the formation of sigmoids. Using the H$\alpha$ data provided by the Multi-channel Subtractive Double Pass spectrograph, \citet{2004SoPh..223..119S} found that the field lines reconnect to form a longer segment, with both the EUV brightenings and flux cancellation being observed, consistent with the model proposed by \citet{1989ApJ...343..971V}. Although proposed for the formation of filaments, this model can interpret  the formation of sigmoids as well. By studying the formation of the helical configuration and associated filament, \citet{2014ApJ...795....4J} also emphasized the importance of magnetic reconnection (also see \citealt{2016ApJ...832...23Y,2016ApJ...829...24V}). \citet{2014ApJ...789...93C} studied the formation of the hot channel-like magnetic flux rope and further confirmed that the twisted field lines, indicated by the continuously sigmoidal hot threads, are formed via the reconnection of two groups of sheared arcades near the PIL. \citet{2017ApJ...834...42J} also analyzed the formation process of a sigmoid and found a transient brightening at the cross site of two J-shaped loops, further supporting the occurrence of the reconnection between the two J-shaped loops.

Nevertheless, the sigmoids previously studied are mostly from regions where the background magnetic field is closed. Whether the property of the background field influences the specific formation and eruption of the sigmoids has been less addressed. In this paper, we investigate the formation and eruption process of a mini-sigmoid originating in a cross-equatorial coronal hole, the magnetic field in which is mostly open, significantly different from that of active regions; thus, it may influence the formation and eruption of the mini-sigmoid. In Section \ref{section2}, we present the observations and the DEM analyses, which are followed by the three-dimensional magnetic field structure and corresponding topology of the mini-sigmoid in Section \ref{section3}. Our conclusion and discussions are given in Section \ref{section4}.

\begin{figure}
     \centering
     \includegraphics[width=6in]{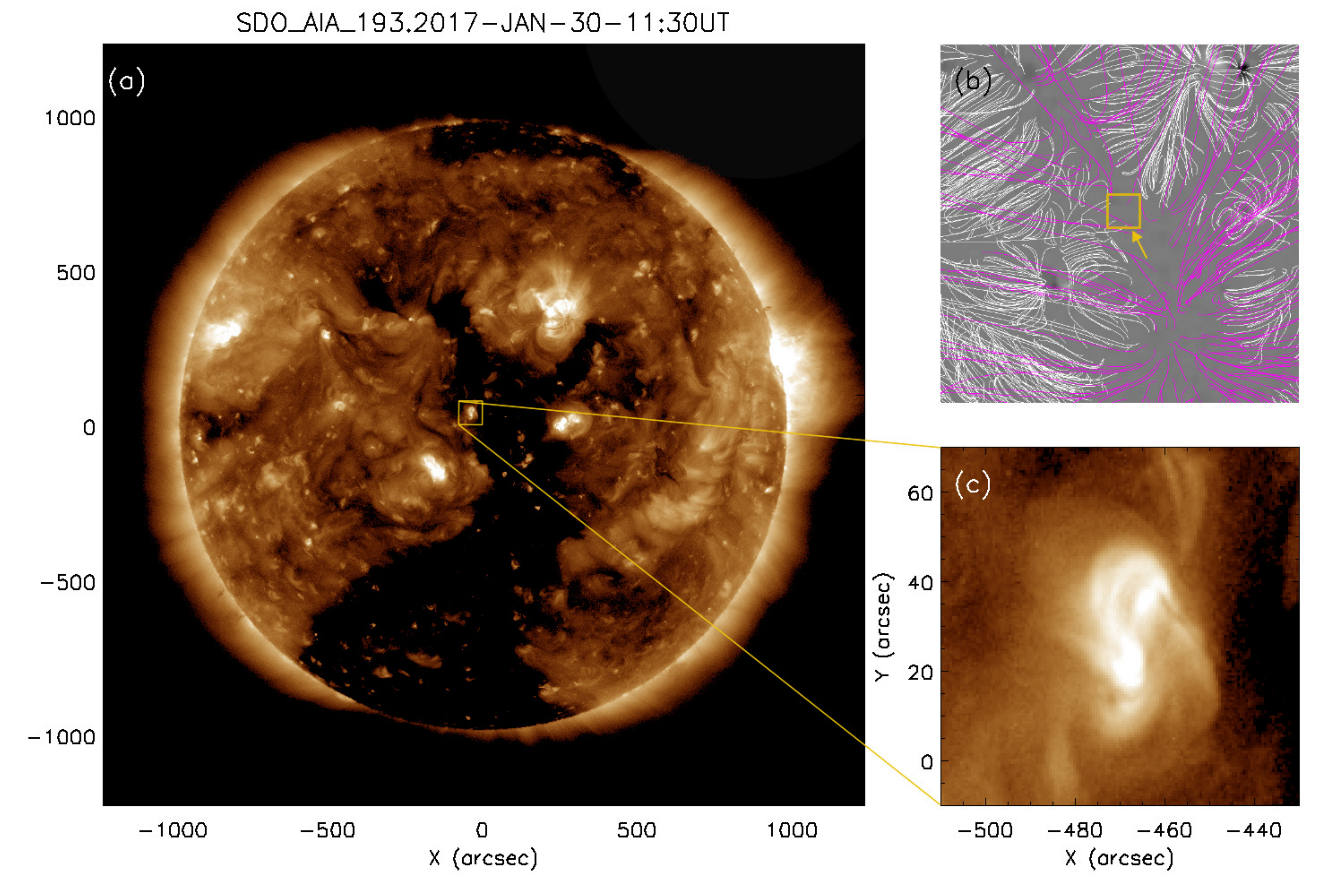}
     \caption{(a) Full-disk image of SDO/AIA 193 Å at 11:30 UT on 2017 January 30; (b) 3-dimensional magnetic field structure computed by the PFSS model, with magenta (white) lines showing the open (closed) field and the yellow arrow pointing to the source region of the mini-sigmoid, same as the region in (a); (c) Zooming in of the mini-sigmoid.}
     \label{2.1}	
\end{figure}

\section{Observations}
\label{section2}
\subsection{Instruments}
The data we utilize are mainly from the Atmospheric Imaging Assembly (AIA; \citealt{2012SoPh..275...17L}) and Helioseismic and Magnetic Imager (HMI; \citealt{2012SoPh..275..229S}) on board \textit{Solar Dynamics Observatory} (\textit{SDO}; \citealt{2012SoPh..275....3P}). The AIA images the solar atmosphere through 10 passbands, effectively covering the temperature that ranges from 0.06 MK to 20 MK, with a temporal cadence of 12 s and a spatial resolution of 1.2 arcsec. The HMI provides the full disk line-of-sight magnetograms, with a cadence of 45 s and a spatial resolution of 1.0 arcsec.

\begin{figure}
     \centering
     \includegraphics[width=7in,trim=20mm 0mm -15mm 0mm,clip]{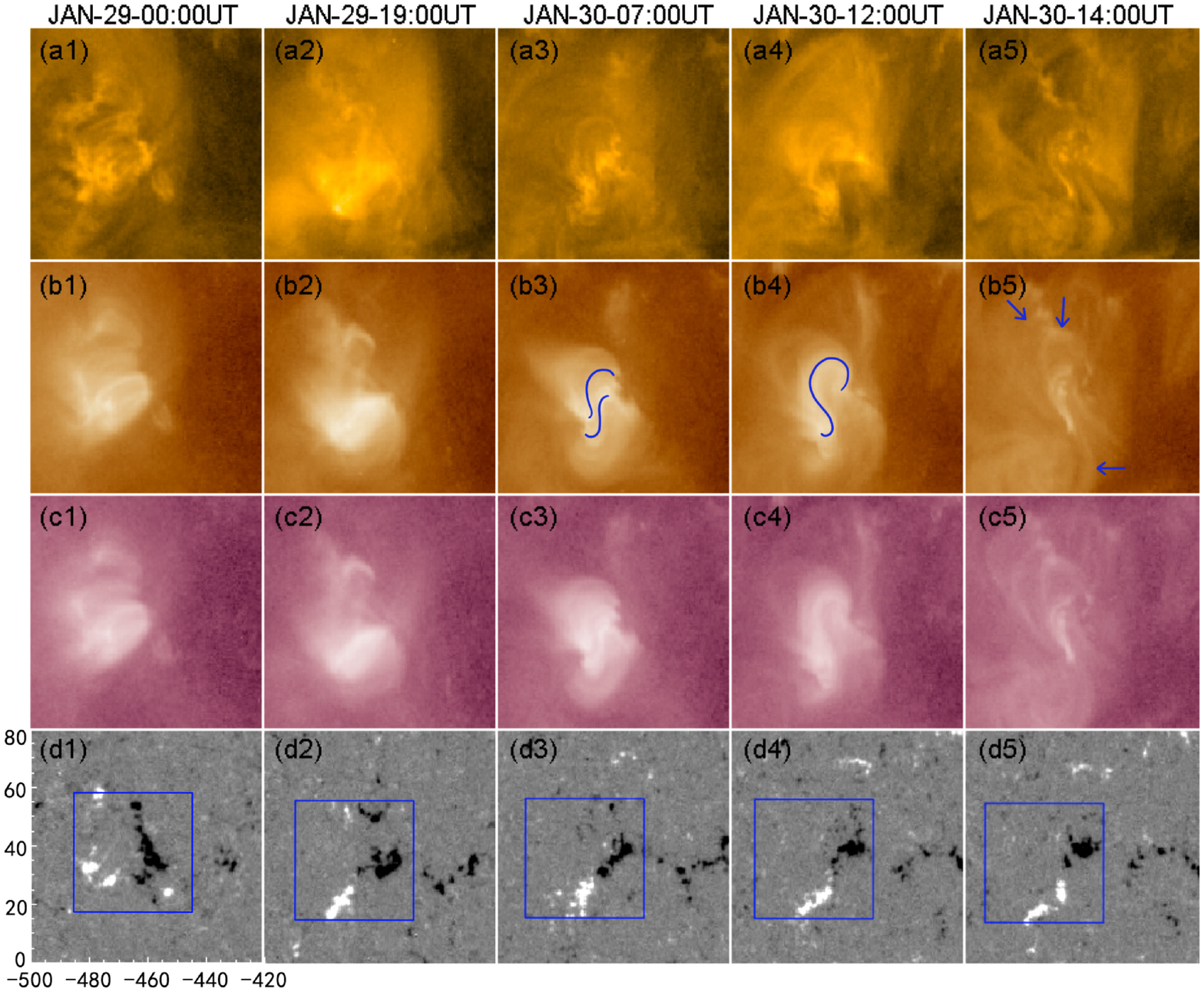}
     \caption{\textit{SDO}/AIA 171 {\AA}, 193 {\AA} and 211 {\AA} images (first three rows) showing the evolution of the mini-sigmoid. The HMI line-of-sight magnetograms (the last row) showing the evolution of the corresponding magnetic field in the photosphere. The two representative J-shaped loops and S-shaped structure are marked by blue lines in panels (b3) and (b4), respectively. The blue arrows in panel (b5) point out the erupting structures. The blue boxes in panels (d1) - (d5) show the region we use to integrate the total positive and negative magnetic flux.}
     \label{2.2}	
\end{figure}

\subsection{Observational Results}
The AIA captured the detailed formation and eruption process of the mini-sigmoid in the coronal hole from 12:00 UT on 2017 January 28 to 11:40 UT on 2017 January 31 (Figure \ref{2.1} and attached movies). The potential field over the source region is shown in Figure \ref{2.1} (b), which is calculated using the potential field source surface (PFSS) model in a spherical volume between r=1 $R_{s}$ and r=2.5 $R_{s}$ \citep{2003SoPh..212..165S}. Here, we directly acquire the data using the ``pfss$\_$viewer.pro" routine in SSW package. The target structure initially appeared as a set of magnetic arcades which were clearly observed at the 171 {\AA}, 193 {\AA}, and 211 {\AA} passbands (Figure \ref{2.2} (a1)--(c1)). With the time going on, these magnetic arcades became brighter (Figure \ref{2.2} (a2)--(c2)). And interestingly, at 07:00 UT on January 30, the magnetic arcades started to shear and became two J-shaped structures with their straight ends being close to each other as shown, in particular, in the 193 {\AA} and 211 {\AA} images (Figure \ref{2.2} (b3) and (c3)). With the development of the shearing, a continuously sigmoidal structure started to appear, as the one shown at 12:00 UT on January 30 (Figure \ref{2.2} (b4) and (c4)). Two representative J-shaped loops transforming into a continuous sigmoidal structure are indicated in Figure \ref{2.2} (b3) and (b4). At 14:00 UT on January 30, most of the bright sigmoid structures erupted, but several small sheared arcades still existed after the eruption (Figure \ref{2.2} (a5)--(c5)). They apparently also make up a diffuse sigmoidal structure as shown in Figure \ref{2.2} (b5).

\begin{figure}
     \centering
     \includegraphics[width=6in]{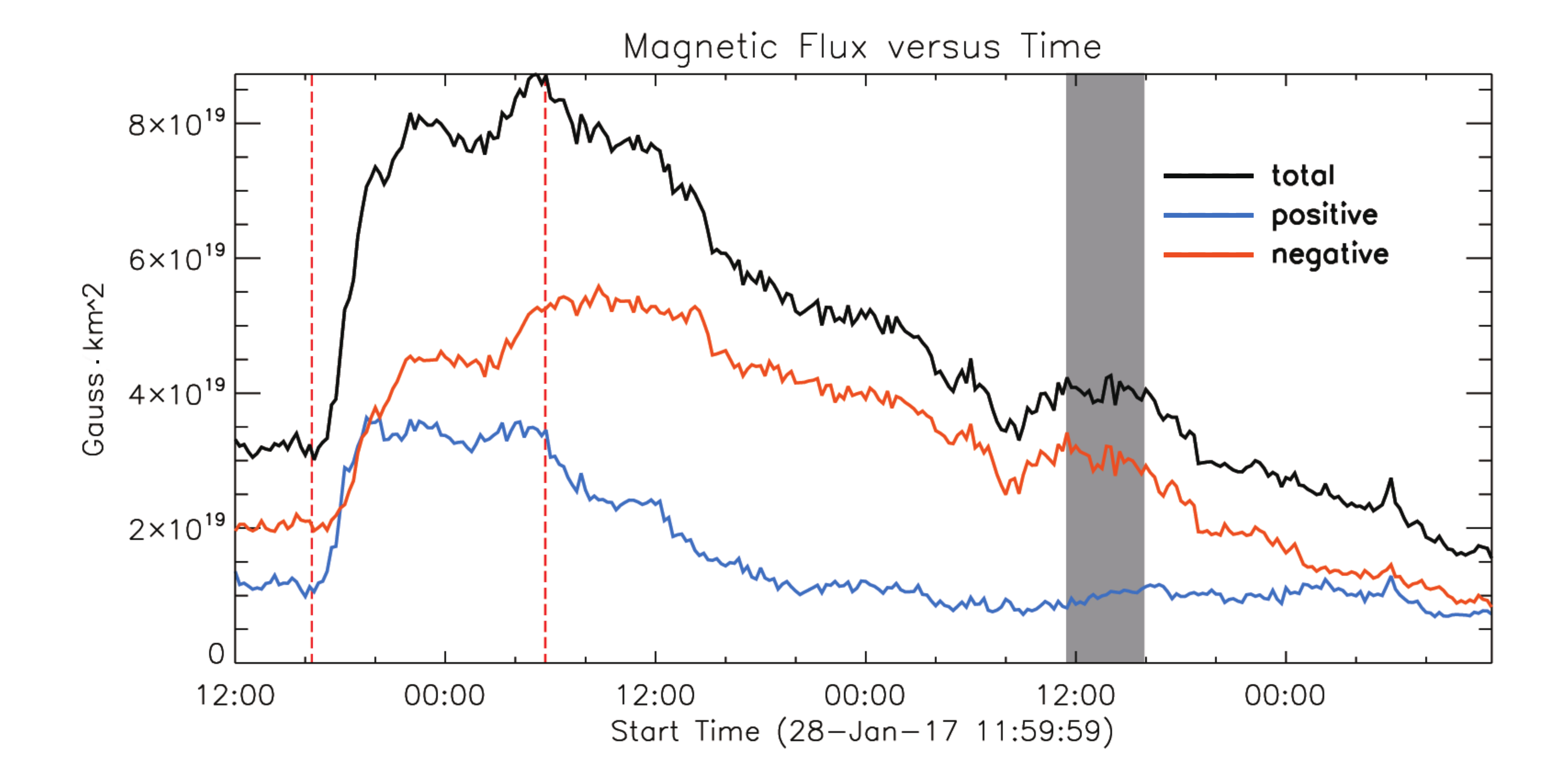}
     \caption{Temporal evolution of the positive (blue), negative (red) and unsigned total magnetic fluxes (black) in the mini-sigmoid source region. The first (second) red dashed line shows the beginning of the emergence (cancellation) phase. The shaded region indicates the time range of the eruption.}
     \label{2.3}	
     \includegraphics[width=6in]{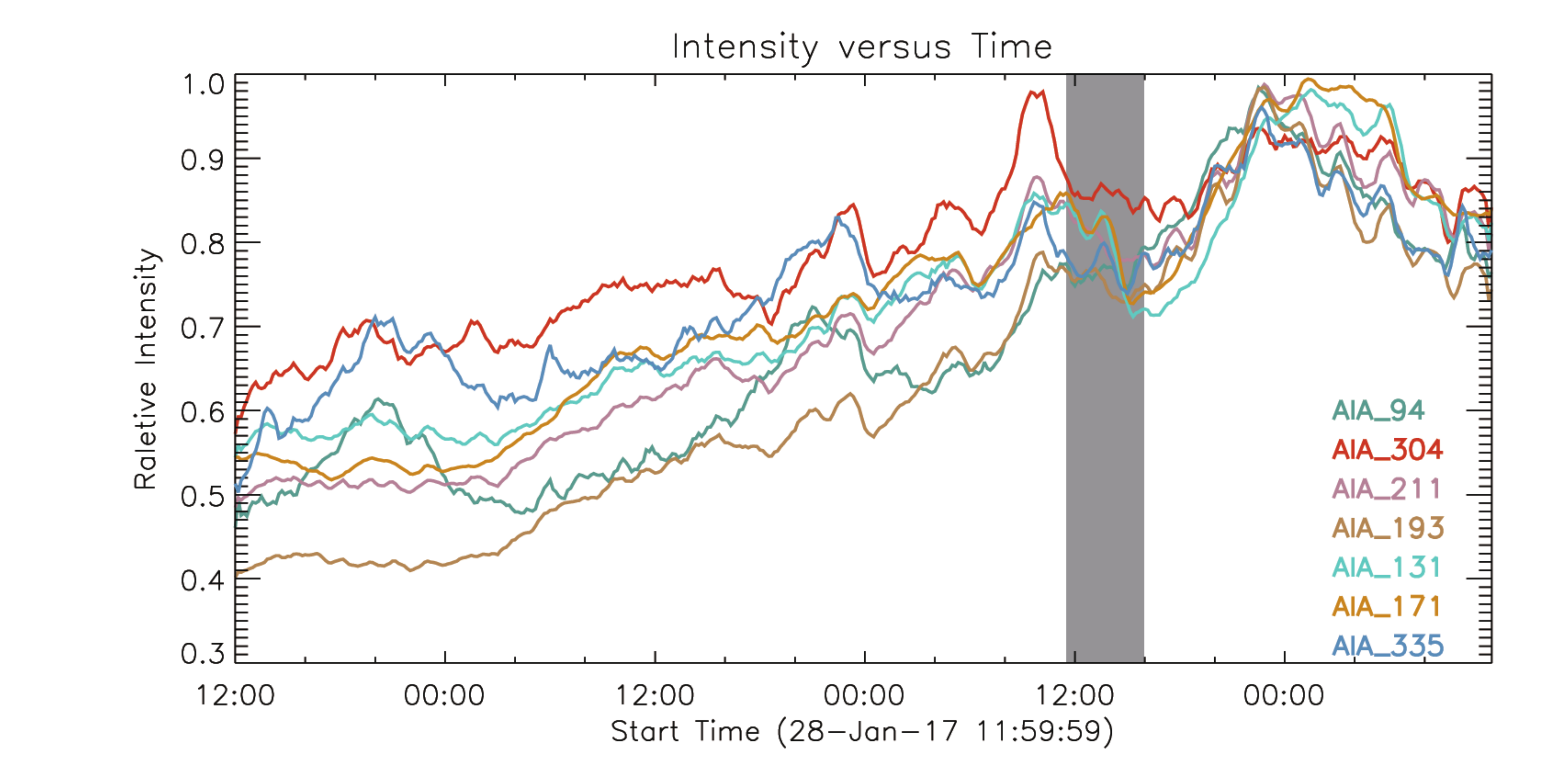}
     \caption{Temporal evolution of the EUV intensities integrated in the whole sigmoid source region at 7 AIA EUV passbands. The shaded region has the same meaning to that in Figure \ref{2.3}.}
     \label{2.4}	
\end{figure}

To understand the formation of the mini-sigmoid, we study the evolution of the HMI line-of-sight magnetic field and find that it can be divided into three stages: a rapid emergence phase, a relatively stable phase that lasted for 8--10 hr, and a long but slow cancellation phase, which can be seen obviously in Figure \ref{2.3}. The total positive and negative fluxes are integrated over the region that moves slightly with time for the sake of including the entire bipole as completely as possible. The moving region, defined according to the magnetic flux centroid of the bipole, is shown in Figure \ref{2.2} (d1) - (d5). One can see that, at about 16:30 UT on January 28, a small bipole began to emerge rapidly with its two polarities separating from each other (Figure \ref{2.2} (d1)). The fast emergence led to an increase in both positive and negative magnetic fluxes (Figure \ref{2.3}). Several hours later, the positive (negative) polarity began to move toward the southwest (northeast) as a result of the shearing motion. Afterwards, they tended to converge to the PIL. Due to the shearing and converging motions, the flux was cancelled there slowly, resulting in a gradual decrease of the positive (negative) flux in the whole area. The flux cancellation is generally thought to be related to the reconnection in the lower atmosphere \citep{1989ApJ...343..971V}, which, in this event,  could change the sheared arcades into the sigmoidal structure, as shown in the AIA images. 

In order to study the temporal variation of the intensity in the sigmoid source region, we plot the integrated EUV flux in the region of interest as shown in Figure \ref{2.4}. One can see that, especially in the flux cancellation phase, the EUV intensities at all passbands increased with time till the occurrence of the eruption. After that, the intensities decreased obviously and then started to increase again. The simultaneity between the increase of the EUV intensity and the decrease of magnetic flux strongly suggests that the reconnection, manifesting as the flux cancellation, could play a key role in the process of the formation of the mini-sigmoid. The reconnection not only transfers the sheared arcades into the sigmoidal field but also heats the surrounding plasma.   

\begin{figure}
    \centering 
    \includegraphics[width=6in]{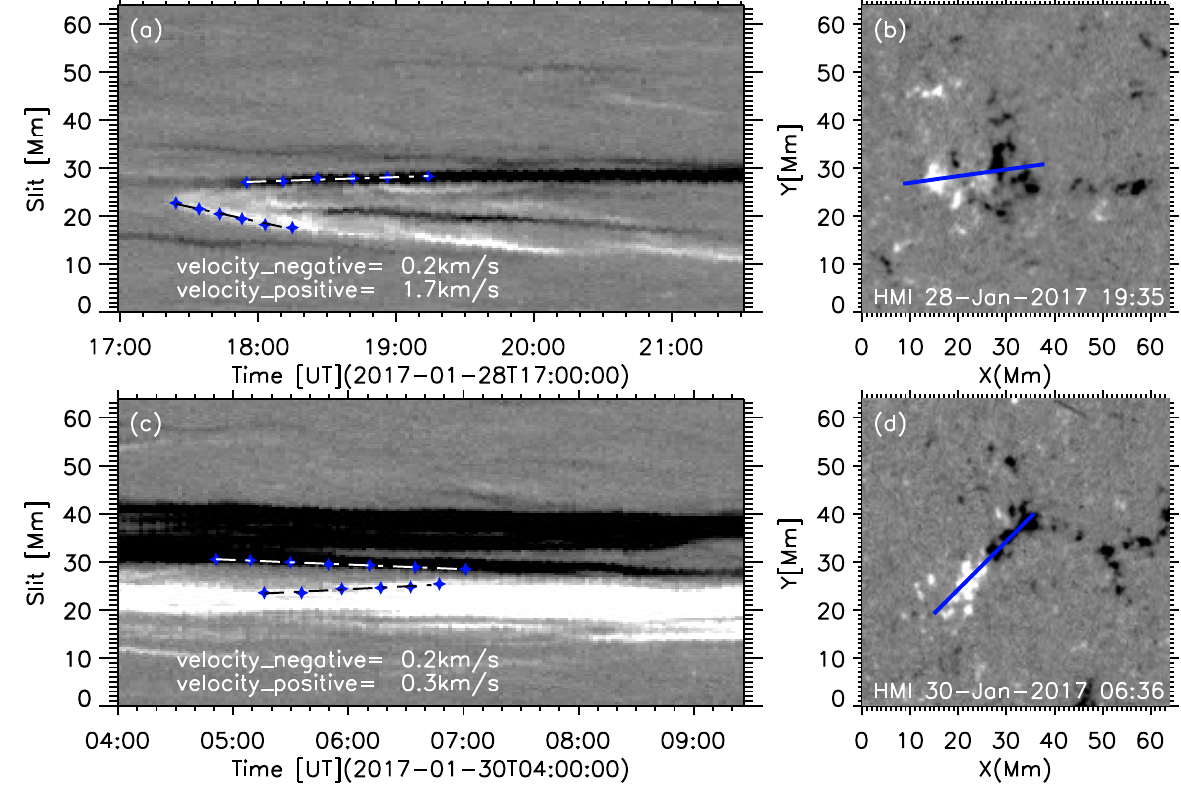}
    \caption{Time-slice plots of HMI line-of-sight magnetograms showing the emergence phase (a) and cancellation phase (c) of the bipolar region. The directions used for making the time-slice plots are indicated by the blue lines in  the two magnetograms at 19:35 UT on 2017 January 28 (b) and 06:36 UT on 2017 January 30 (d). The slopes of the oblique lines represent the moving velocities of the positive and negative polarities.} 
    \label{2.5}
\end{figure}

\begin{figure}
    \centering 
    \includegraphics[width=6in]{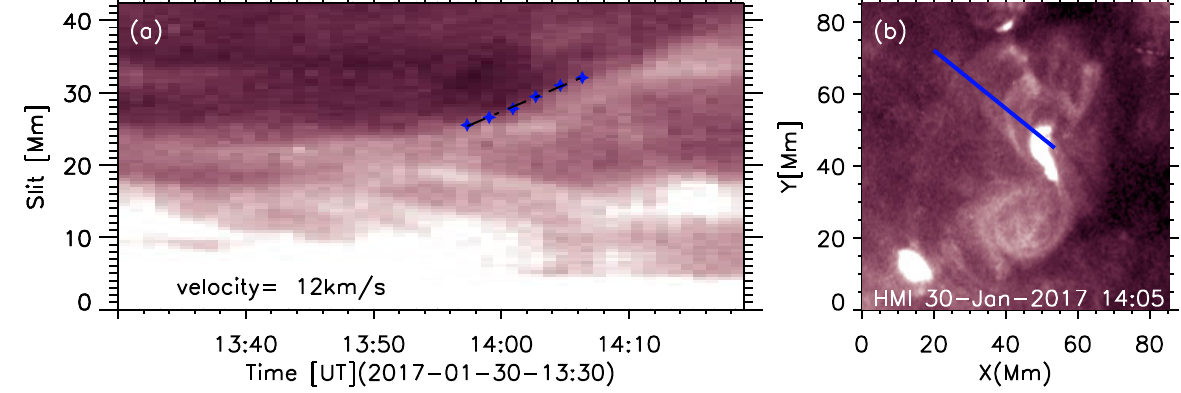}
    \caption{(a) Time-slice plot of AIA 211 {\AA} images showing the eruption process. (b) The slice is taken along the direction to the northwest (the line in blue) as indicated in the AIA 211 {\AA} image at 14:05 UT on 2017 January 30.  The slope of the oblique line in panel (a) represents the moving velocity of the erupting structure along the plane of sky.} 
    \label{2.6}
\end{figure}

To further quantify the movement of the positive and negative polarities, we make time-slice diagrams of the HMI images as shown in Figure \ref{2.5}, where the panels (a) and (c) show the emergence phase and the flux cancellation phase, respectively. The directions of the slices we choose are shown in Figure \ref{2.5} (b) and (d). We find that the velocity of the positive polarity (1.7 km s$^{-1}$) is much larger than that of the negative one (0.2 km s$^{-1}$) in the emergence phase. In the short period of 04:00–09:00 UT on January 30, during which the flux cancellation can be clearly detected, the convergence velocities of the positive and negative polarities are estimated to be about 0.3 km s$^{-1}$ and 0.2 km s$^{-1}$, respectively. The time-slice plot of AIA 211 {\AA} images show the dynamic process of the eruption of the sigmoid (Figure \ref{2.6}). Unfortunately, restricted by the projection effect, we cannot determine the real height of the erupting structure and its evolution with time, but can only derive its projection distance from the source region as shown in Figure \ref{2.6} (b) and the velocity component in the plane of sky in Figure \ref{2.6} (a). Assuming that the eruption is along the radial direction, the linear velocity is corrected to be around 60 km s$^{-1}$.

\begin{figure}
     \centering
     \includegraphics[width=6.8in,trim=5mm 0mm -5mm 0mm,clip]{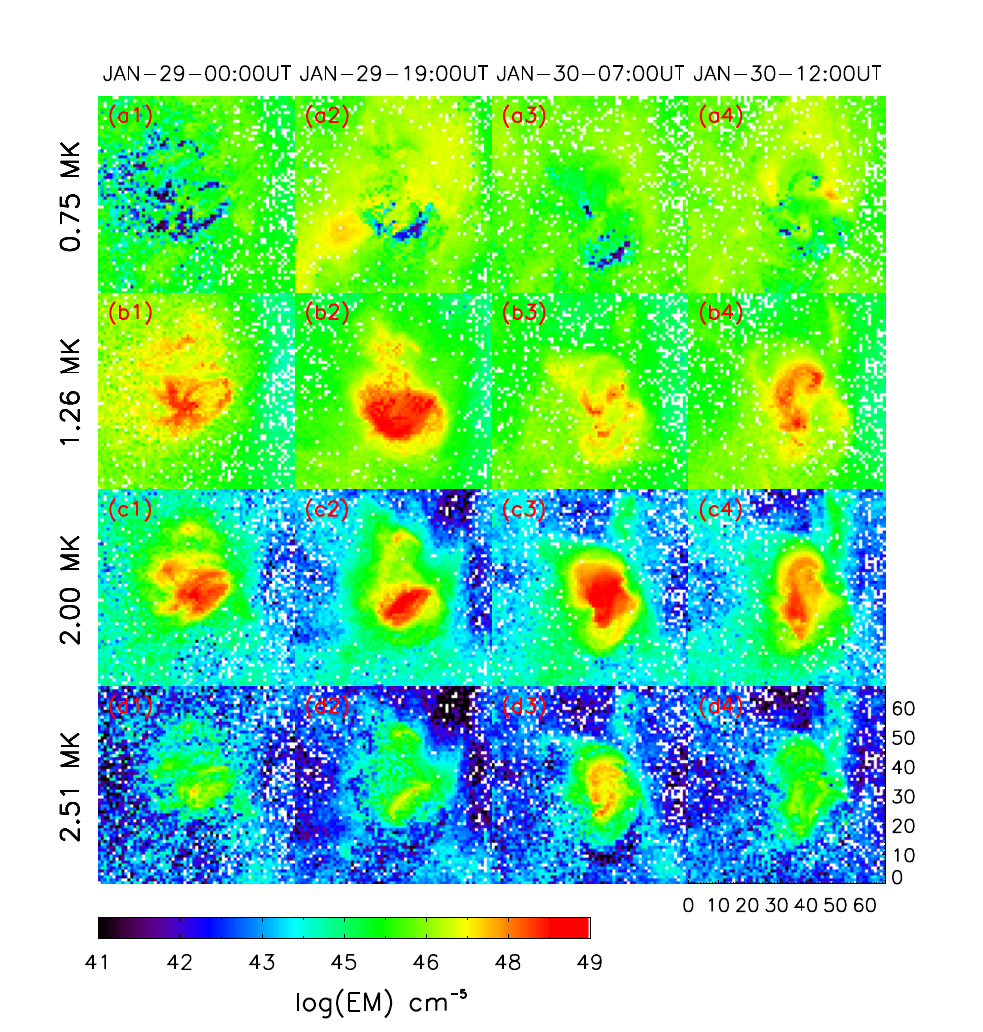}
     \caption{Temporal evolution of the EM distributions of the sigmoid source region at the four temperatures intervals ranging from 80 MK to 250 MK .}
     \label{2.8}	 
\end{figure}

\subsection{Thermal Property of Mini-sigmoid}

The observed flux of passband $i$ can be written as $F_i=\int R_i(T) \times DEM(T) dT$, where the $R_i(T)$ is the temperature response function and $DEM(T)$ is the plasma DEM in the corona. Knowing the observed flux $F_i$ and response function $R_i(T)$, we resolve the DEMs of each pixel in the mini-sigmoid source region following the method in \citet{2012ApJ...761...62C}, using the ``xrt$\_$dem$\_$iterative2.pro'' routine in SSW package. The emission measure maps at the four temperatures are shown in Figure \ref{2.8}. One can see that the temperature of the sigmoidal structure is around 2.0$\times10^{6}$ K (Figure \ref{2.8} (b4) and (c4)). It is interesting that, before the appearance of the sigmoid, two J-shaped structures that constitute an apparent sigmoid have a higher temperature of $2.5\times10^{6}$ K (Figure \ref{2.8} (d3)). Such a difference in temperature between the double J-shaped loops and the S-shaped structure is in agreement with the expectation that the sigmoidal field lines, are formed by the reconnection between two sheared arcades near the PIL. With the J-shaped arcades being closer to the reconnection site, they tend to be heated up to a higher temperature. Shortly afterwards, these J-shaped arcades transform into the sigmoidal structures, which then cool down probably due to the expansion. This is basically consistent with the observations by \citet{2009ApJ...698L..27T}.

\section{Nonlinear Force-free Field Structures of Mini-sigmoid}
\label{section3}
\subsection{Flux-rope Insertion Method}
Through the flux-rope insertion method \citep{2004ApJ...612..519V,2008ApJ...672.1209B,2009ApJ...704..341S}, we further reconstruct the nonlinear force-free field structures of the mini-sigmoid source region using the HMI line-of-sight magnetic field as the bottom boundary. More details for the flux-rope insertion method can be found in \citet{2009ApJ...704..341S}.

In summary, the flux-rope insertion method usually follows four steps: (1) calculating the potential field as the initial condition; (2) creating a cavity and then inserting a flux rope along a selected path; (3) choosing the axial flux and poloidal flux of the inserted rope; (4) evolving the structure to the NLFFF through magneto-frictional relaxation and then comparing them with the observations to achieve the best fitted model. In our calculation, we use the uniform grid along the longitudinal and the latitudinal direction and the exponentially stretched grid along the radial direction, with the size of the bottom row is $dx=dy=1\ cell=0.001$ solar radii and $dz=0.7\ Mm$.

\begin{figure}
     \centering
     \includegraphics[width=6in]{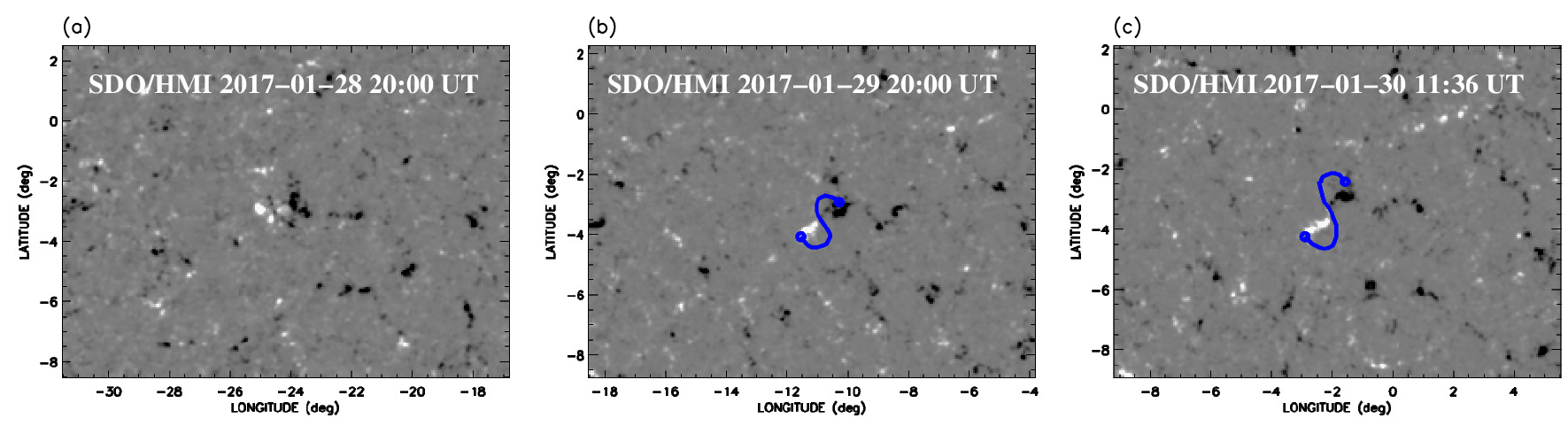}
     \caption{HMI magnetograms showing the radial component of magnetic field in the region of interest. The blue curves in panels (b) and (c) show the paths along which the flux ropes are inserted with the blue circles denoting their footpoints.}
     \label{3.1}	
\end{figure}

Here we choose three time instances, when the target region exhibited clearly recognized feature for comparison: (1) 20:00 UT on January 28, when the emergence phase almost finished, a row of arcades that connected the positive and negative polarities of the bipolar region are clearly observed at the AIA 193 {\AA} passband; (2) 20:00 UT on January 29, when the emerged arcades were severely sheared; (3) 11:32 UT on January 30, when the flux cancellation phase had started and a north-south directed sigmoid had been formed as was clearly observed at the 193 {\AA} passband. The paths of the inserted flux ropes are shown in Figure \ref{3.1}. Moreover, in order to simplify the analysis, the inserted flux ropes only have an axial flux. The paths of the inserted flux ropes are shown in Figure \ref{3.1}.

\begin{figure}
     \centering
     \includegraphics[width=6.5in,trim=5mm 5mm -5mm 0mm,clip]{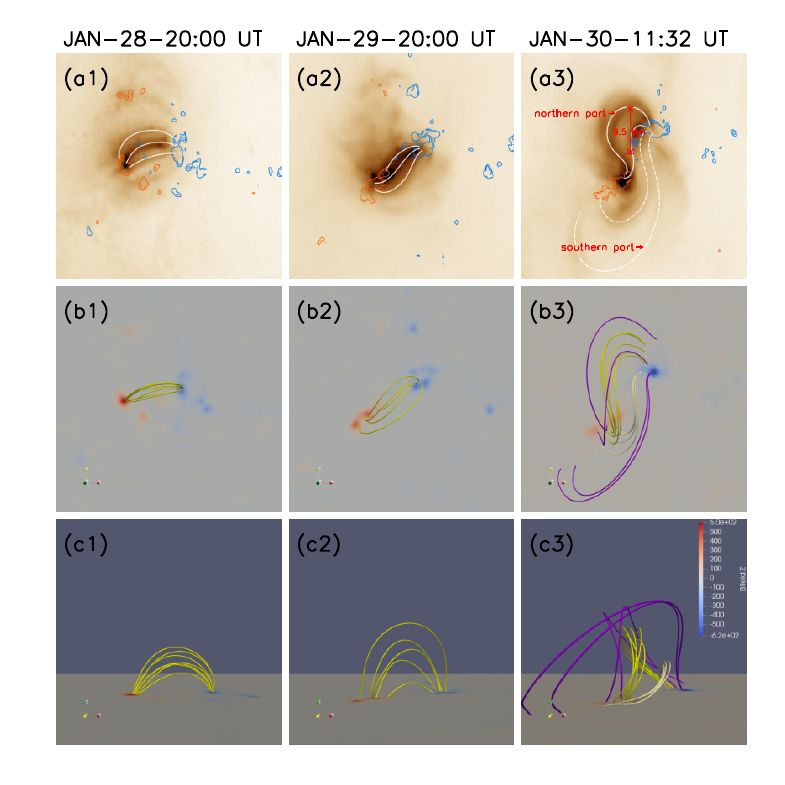}
     \caption{NLFFF structures with the corresponding AIA 193 {\AA} images. (a1)-(a3) AIA 193 {\AA} images overlaid by the contours of the HMI line-of-sight magnetic field. The red (blue) indicates the positive (negative) polarity. The sheared arcades and sigmoidal structures are delineated by the white lines. (b1)-(b3) Top view of the extrapolated field lines. (c1)-(c3) Side view of the sigmoidal lines. The yellow lines show the observed sigmoidal structure, and the purple lines indicate those with the highest twist.}
     \label{3.2}	
\end{figure}

\subsection{NLFFF Structures}
\begin{figure}
     \centering
     \includegraphics[width=6.3in,trim=5mm 0mm -5mm 0mm,clip]{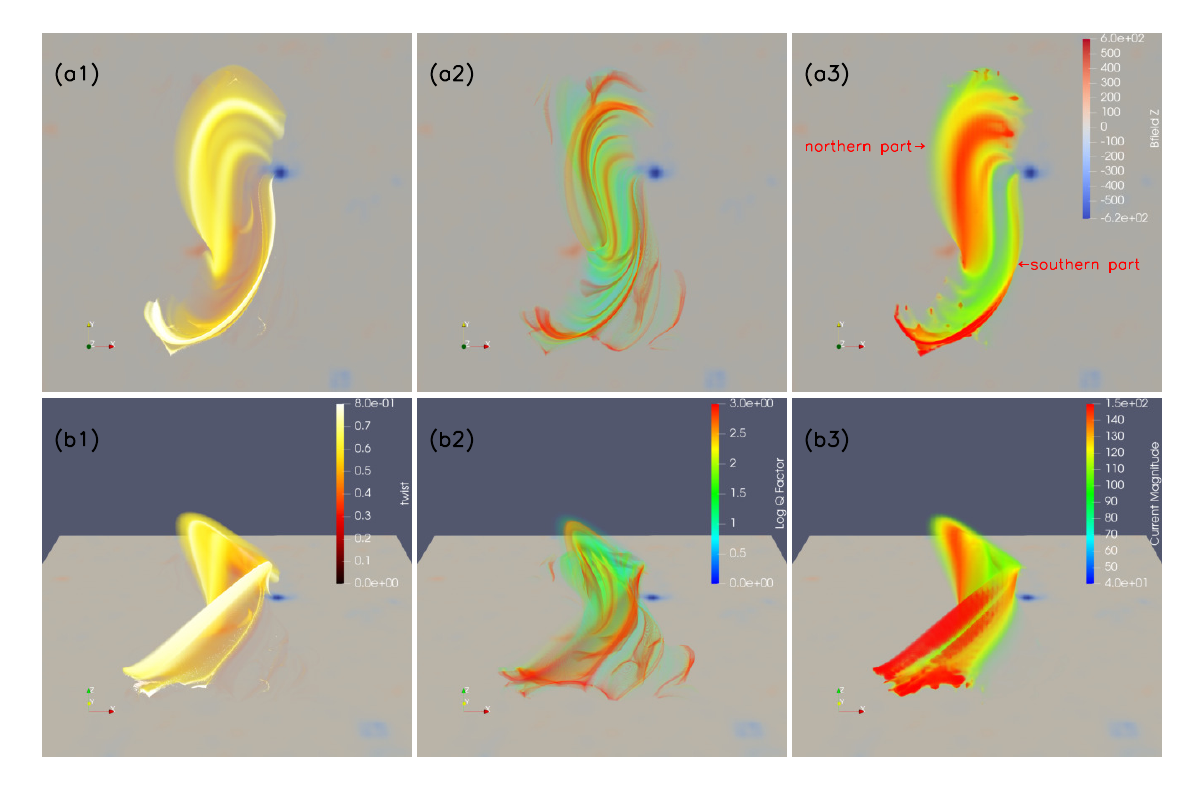}
     \caption{(a1)-(a3) Top view of 3-dimensional distribution of the twist, squashing factor Q and electric current, respectively. (b1)-(b3) Same as panels (a1)-(a3) but for side view. }
     \label{3.4}
\end{figure}

Figure \ref{3.2} shows the NLFFF structures at the three time instances. The overlaid white lines in Figure \ref{3.2} (a1)-(a3) show the observed arcades and sigmoidal structures for comparison with the extrapolated results. Note that the dashed lines in Figure \ref{3.2} (a3) mark very faint sigmoidal loops that can be identified in the corresponding NLFFF. One can see that, the magnetic field lines that are best fitted to the observations (the middle and bottom rows of Figure \ref{3.2}) are gradually sheared and finally form the sigmoidal ones with the increase of the inserted axial flux. Table \ref{b1} shows the axial flux of the inserted ropes as well as the potential field energy, magnetic free energy and relative magnetic helicity of the corresponding NLFFF field. At the first instance, the potential field model matches the observations better than the NLFFF model, which implies that the magnetic field that just emerges is most likely close to potential. At the second instance, the magnetic arcades have been seriously sheared (Figure \ref{3.2} (b2)), which correspond to the bright structure in the source region. At the third instance, the sigmoidal field lines have been formed. We observe a number of S-shaped loops with the southern footpoints distributed at a very extended area, as shown in Figure \ref{3.2} (a3). The northeastern sigmoidal loops were much brighter than the southwestern ones. Both of them can be identified in the extrapolated magnetic fields as denoted by yellow and light-yellow lines in Figure \ref{3.2} (b3), respectively. The purple lines in Figure \ref{3.2} (b3) and (c3) show the strongly twisted field lines corresponding to the high-twist region in Figure \ref{3.4} (a1) and (b1). Compared with the previous instances, the twist of the field lines is significantly increased, consistent with the transformation of sheared arcades into a sigmoidal field. Note that, we check the convergence of the reconstructed fields to the force-free state by calculating the average of the angle between the magnetic field and current density $\sigma_J=\Sigma_i (|\vec J \times \vec B|_i/B_i)/\Sigma_i J_i$ and also the divergence-free state by calculating the average value of the magnitude of $f_i\approx (\nabla \cdot \vec B)_i \Delta V_i/(B_iA_i)$ following \citet{2000ApJ...540.1150W}. It is found that both values, as shown in Table \ref{b2}, are very small, consistent with the assumption of force-free field.
\begin{table}
     \centering 
     \caption{Parameters of the Best-fitting Results}
     \begin{tabular}{ccccc}
     \hline
     \hline
     Time & Axial Flux & Potential Field Energy & Magnetic Free Energy & Helicity \\ 
       & (Mx) & (erg) & (erg) & (Mx$^2$) \\
     \hline 
     01-28 20:00&  & $6.66\times10^{29}$ &  & $1.95\times10^{40}$ \\ 
     01-29 20:00& $2\times10^{19}$ & $7.36\times10^{29}$ & $5.98\times10^{27}$ & $2.43\times10^{40}$ \\
     01-30 11:32& $1\times10^{20}$ & $7.69\times10^{29}$ & $5.37\times10^{28}$ & $3.00\times10^{40}$ \\ 
     \hline
     \end{tabular} 
     \label{b1}
\end{table}

\begin{table}
     \centering 
     \caption{Force-free and Divergence-free Measure}
     \setlength{\tabcolsep}{7mm}{
     \begin{tabular}{llllll}
     \hline
     \hline
     Time(UT) & $\sigma_J$ & $f_i$  \\ 
     \hline 
     January 28 20:00 UT & $1.815\times10^{-2}$ & $1.255\times10^{-3}$ \\ 
     January 29 20:00 UT & $1.678\times10^{-2}$ & $7.880\times10^{-4}$ \\
     January 30 11:32 UT & $2.556\times10^{-2}$ & $5.643\times10^{-4}$  \\ 
     \hline
     \end{tabular}}
     \label{b2}
\end{table}

\subsection{Twist, Squashing Factor and Electric Current}

In order to further quantify the property of the sigmoidal field, we calculate the distribution of the twist at the third instance (shown in Figure \ref{3.4} (a1) and (b1)). It is found that the sigmoidal field has a highest twist of 0.8, corresponding to the erupting threads that distribute in both northern and southern parts of the mini-sigmiod. We calculate the squashing factor $Q$, which measures the change of magnetic field connectivity \citep{1996A&A...308..643D, 2002JGRA..107.1164T, Titov_2007}, following the method of \citet{Tassev_2017} and \citet{Scott_2017}. Similar to what is revealed in observed EUV images, the distribution of the $Q$ value also presents two groups of sigmoidal structures. The northern part shows an obvious S-shape from the top view, corresponding to the brightest sigmoidal loops. It It is argued that the observed sigmoidal loops are most likely associated with magnetic reconnection, which is prone to occur in high-$Q$ region as found in many previous active region events \citep{1997A&A...325..305D, Mandrini1997, Masson_2009, Vemareddy_2014,Yang_2015, Savcheva_2015, 2016A&A...591A.141J} and numerical simulations \citep{2005A&A...444..961A, 2010A&A...516A...5W, 2013A&A...555A..77J}. The current distribution is also found to be S-shaped, consistent with the morphology of the pre-erupting EUV structures. It further implies that the observed sigmoidal structure is closely related to electric current dissipation in the reconnection region. 




From Figure \ref{3.4} (a1) and (b1), one can see that the maximum of the twist is around 0.8. This is to say, a full flux rope in the coronal hole-hosted mini-sigmoid has not been formed when approaching the eruption. It shows that the mini-sigmoid mostly consists of strongly sheared loops. Considering the low twist of the sigmoid, the occurrence of kink instability is impossible. This is also consistent with the absence of the rotation motion during the eruption, which is usually used to identify the appearance of kink instability (e.g., \citealt{2003ApJ...595L.135J, Williams_2005, Rust_2005}).

\begin{figure}
     \centering
     \includegraphics[width=4.5in]{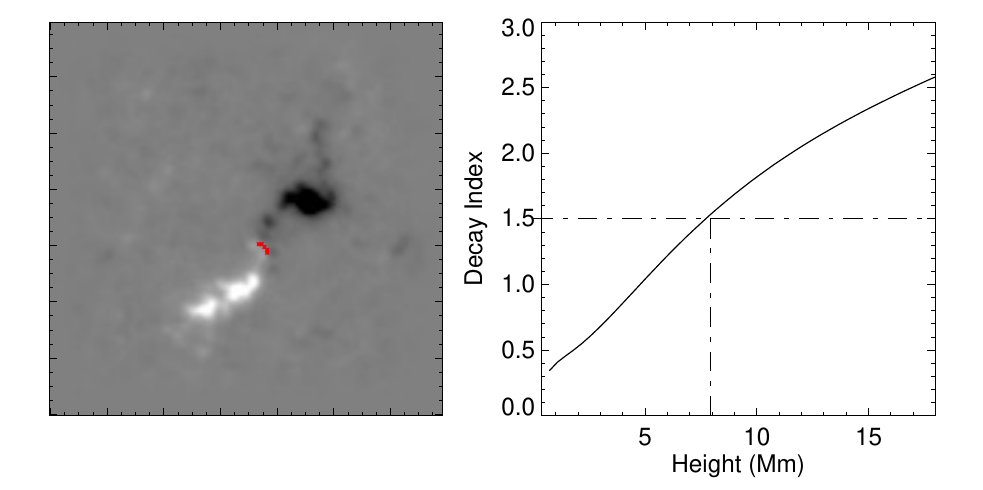}
     \caption{HMI line-of-sight magnetogram at 11:32 UT on January 30 (left panel) and the distribution of the decay index at a place above the PIL (in red) with height (right panel). The horizontal dash-dotted line shows the decay index of 1.5, the corresponding height is indicated by the vertical dash-dotted line.}
     \label{3.5}	
\end{figure}

Different from the sigmoids in active regions, the mini-sigmoid under study is located inside a coronal hole, where the background magnetic field is mostly open and thus more favorable for the eruption of the mini-sigmoid. Here, we calculate the decay index of the background field above the PIL before the eruption using the equation $n=-d lnB/d lnh$, where $B$ and $h$ denote the background field strength and height, respectively. The background field is computed by means of the potential model based on the magnetogram took at 16:30 UT on January 28. The reason of selecting the magnetogram even before the emerging phase is mainly for eliminating the contribution of the bipole to the background field. Note that only the horizontal component of the potential field is used to calculate the decay index, since the vertical component does not constrain the eruption \citep{Cheng_2011,Nindos_2012}.  The distributions of the background field and corresponding decay index with height are shown in Figure \ref{3.5}. 

From the reconstructed 3-dimensional sigmoidal structure, we estimate that the mini-sigmoid has ascended to a height of more than 8 Mm at the third instance where the decay index is found to be 1.5, which is very close to the critical value of the torus instability (e.g., \citealt{1978mit..book.....B, PhysRevLett.96.255002}). The real height of the erupting sigmoidal structure is hard to be determined because it is located near the center of the solar disk. However, we estimate the distance of the northern part of the sigmoid to the PIL, which is about 9.5 Mm. From Figure \ref{3.5}, we can see that the corresponding decay index is about 1.7 if assuming this distance is approximately the sigmoid height. Note that such a value is a lower limit, the real one should be bigger as the real height could be larger than 9.5 Mm. It shows that the open background field declines very quickly with height so that the torus instability can start in the very low corona, thus in favor of the eruption. On the other hand, it may also have a role in preventing the twist from being continuously accumulated to a higher value. In fact, a recent simulation by \citet{Fan_2017} also stated that the rising flux can erupt before the twist increases to 1 once it reaches the critical height of torus instability. In short, we argue that the mini-sigmoid may have entered the unstable regime at the third instance, which is consistent with the following eruption process as seen in the AIA images.


\section{Summary and Discussions}
\label{section4}



In this study, we explore the formation and eruption of a mini-sigmoid located within a cross-equatorial coronal hole, where the magnetic field is mostly open and significantly different from that of sigmoids in active regions as studied previously. Through analyzing the EUV images from \textit{SDO}/AIA and the magnetic field data from \textit{SDO}/HMI, we find that the source region of the mini-sigmoid initially appears as a bipolar region. Subsequently, it experiences three phases: a rapid emergence phase, a relatively stable phase, and a long but slow cancellation phase. In particular, shortly after the emergence phase, the positive and negative polarities become sheared to each other. Such a shearing motion can drive the evolution of the magnetic field from near-potential to non-potential \citep{2010ApJ...708..314A}. Thereafter, the converging motion makes the two ends of the non-potential sheared arcades with opposite directions reconnect and build up the sigmoidal field lines. The continuous flux cancellation near the PIL indicates that the reconnection takes place in the lower atmosphere. Moreover, in the EM maps of different temperatures, we see two J-shaped structures with a higher temperature, which then transform into a sigmoidal structure. The temperature of the sigmoidal field slightly decreases after its formation but is still larger than that of the background coronal hole. This indicates that the reconnection also plays an important role in heating the local plasma.  


We also reconstruct the potential field and nonlinear force-free field structures of the mini-sigmoid source region using the flux rope insertion method. The results well reproduce the potential magnetic arcades, non-potential sheared arcades, and their transformation into the sigmoidal structure. The distribution of the squashing factor $Q$ further reveals that the S-shaped field lines cross the high-Q region, suggesting that the reconnection is most likely to occur there.

It has been realized that the sigmoids notably vary in scale \citep{2014SoPh..289.3297S}. Interestingly, it has also been found that the small-scale sigmoids have the same formation and eruption mechanism regardless of their scale \citep{Liu_2002, 2005AdSpR..36.1579M, Jiang_771, Jiang_780, Chesny_2015, Chesny_2016, Vasantharaju_2019}. It is worth noticing that all the small-scale or mini-scale sigmoids that have been analyzed previously originated from active regions or weak field regions, where the background field is mainly closed. However, the mini-sigmoid in the present study was located within a coronal hole. On the one hand, similar to previous studies, the formation of the mini-sigmoid is comparable with that of the sigmoids in active regions, basically conforming to the model proposed by \citet{1989ApJ...343..971V}. It requires continuous shearing and converging motions near the touch point, two J-shaped loops then form continuous S-shaped structures through the tether-cutting like reconnection \citep{2009ApJ...700L..83G,2010ApJ...725L..84L,2014ApJ...789...93C,2015ApJ...804...82C}. As \citet{2004A&A...413L..23K} suggested, the sigmoidal structure denotes the field lines coming from the reconnection region below the rising rope, where the plasma are heated correspondingly. On the other hand, it may be easier for the mini-sigmoid in the coronal hole to erupt due to the open background field. This is supported by the fact that the decline of the background field above the mini-sigmoid is fast enough at the height below 10 Mm. The different background field may also influence the critical twist before the eruption. For active-region sigmoids, the twist is often found to be more than 1 (e.g. \citealt{Romano2003, 2009ApJ...703.1766S, Chen_2014, 2018ApJ...855...77S}), which is much larger than that (0.8) of the mini-sigmoid in the current study. This is understandable as the closed background field provides stronger magnetic tension comparing with the open one, thus in favor of the accumulation of the twist. Note that, in coronal hole, coronal jets and whip-like motions are often observed \citep{2007Sci...318.1580C}. They are believed to be caused by the reconnection of emerging flux with the background open field \citep{1992PASJ...44L.173S, 2000eaa..bookE2272S}, thus possibly different from the sigmoid eruption studied here.


It is worth noting that, the formation of the mini-sigmoids involves a much smaller magnetic flux than the active region sigmoids. \citet{2011A&A...526A...2G} studied the formation of the sigmoid in NOAA AR 10977 and found that a magnetic flux of about $7\times10^{20}$ Mx was cancelled in the source region during the period of three days. \citet{2012ApJ...759..105S} also found a flux of about $10^{21}$ Mx being cancelled during the decay phase. The cancelled flux for the mini-sigmoid in this work is only $5\times10^{15}$ Mx (positive) / $9\times10^{15}$ Mx (negative). Correspondingly, the total helicity of the active-region sigmoids is more than $1\times10^{41}$ Mx$^{2}$ \citep{2011A&A...526A...2G,2012ApJ...759..105S}, while that of the small-scale one we study is only about 1--3$\times10^{40}$ Mx$^{2}$. Nevertheless, although small amount, the cancelled flux during the formation of the mini-sigmoid amounts for more than 70{\%} of the peak flux, much larger than the values (34{\%}) found in the active region cases \citep{2011A&A...526A...2G,2012ApJ...759..105S}. 


\acknowledgements  

We thank Yang Guo, Kai E. Yang, Ze Zhong for valuable discussions, and we thank the team of SDO/AIA and SDO/HMI for providing the valuable data. AIA data and HMI data are courtesy of NASA/SDO, which is a mission of NASA’s Living With a Star Program. Z.W.H., X.C. and M.D.D. are funded by NSFC grants 11722325, 11733003, 11790303, 11790300, 11961131002, Jiangsu NSF grants BK20170011, sand “Dengfeng B” program of Nanjing University. Y. N. Su and T. Liu are supported by NSFC 11473071, 11790302 (11790300), 41761134088.



\newpage
\pagenumbering{Roman}


\clearpage

\end{document}